# Title: Observation of Order and Disorder in Solid-Electrolyte Interphases of Lithium-Metal Anodes

**Authors:** Hyeongjun Koh[a,#], Eric Detsi[a], Eric A. Stach[a,b*]

**Affiliations:**
[a] Department of Materials Science & Engineering, University of Pennsylvania, Philadelphia, PA 19104, USA
[b] Laboratory for Research on the Structure of Matter, University of Pennsylvania, Philadelphia, PA 19104, USA
[#] *Present address: Andlinger Center for Energy and Environment, Princeton University, New Jersey, NJ 08540, USA*

*Corresponding author and lead contact: stach@seas.upenn.edu

**Abstract:**
Battery interfaces critically influence lithium-metal battery performance through their role in ion diffusion and dendrite formation. However, structural characterization of these interfaces has remained challenging due to limitations in high-resolution methods and artifacts from electron irradiation. Using cryogenic conditions for both specimen preparation and scanning electron nanobeam diffraction, we can determine the structural organization at the interface between the vitrified electrolyte and adjacent layers. We identified two distinct interface types: one showing short-range order adjacent to lithium metal, and another displaying a mixed structure of short-range ordering and defective lithium fluoride nanoscale crystallites at a copper collector. Notably, short-range order appeared exclusively in electrolytes demonstrating high reversibility. Our results establish that solid-electrolyte-interphase structure directly influences lithium deposition morphology and battery performance. This methodology opens new possibilities for high-resolution characterization of interfaces in energy storage materials, advancing our understanding of their critical structural properties.

**Abstract Graphic**

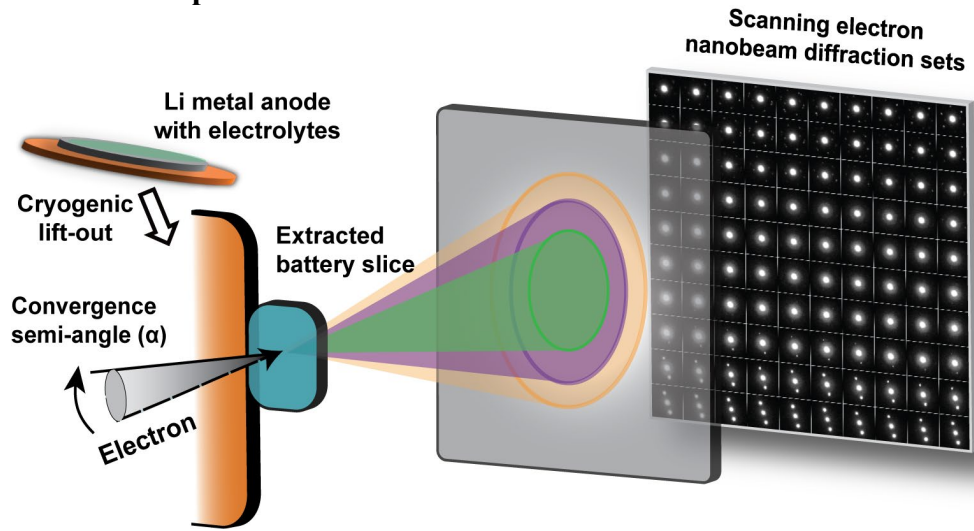

**Main Text**

In rechargeable lithium-metal batteries, the liquid electrolyte-metal electrode interface requires ion conductivity and electrical insulation for efficient performance. The mechanical stability of this interface is also crucial for suppressing dendrite formation[1–3]. The structural properties of this interface control both ion movement across the boundary and how uniformly metal is deposited during charging, which directly impacts the battery's safety and longevity[4–6]. However, imaging these nanoscale interfaces remains challenging, as conventional sample preparation and imaging methods can introduce artifacts[7–9].

Phase contrast imaging, commonly employed in high-resolution transmission electron microscopy (HRTEM), requires precise focusing and careful operation, yet it often leads to significant electron beam-induced damage[10,11]. This approach has been recently used in observing the solid electrolyte interphases (SEIs), likely altering their true structures. To overcome these limitations, we utilize cryogenic electron microscopy (cryo-EM) techniques, including scanning electron nanobeam diffraction (SEND) and electron energy loss spectroscopy (EELS), to characterize two distinct interface types in lithium-metal batteries and their correlation with battery performance. By employing low-dose, pixelated diffraction patterns[12], this approach minimizes electron irradiation damage and enables damage-free characterization of SEI. These insights provide a new framework for designing more efficient energy storage systems and demonstrate the power of cryogenic imaging for studying sensitive material interfaces.

**Figure 1a** shows a schematic of a battery extracted from a Li-metal coin cell used in our experiments (see Methods in Supporting Information for details). The extraction was performed by cryogenic lift-out to reveal embedded interfaces underneath vitrified liquid electrolytes. The extracted Li metal slices were welded to a commercial TEM half-grid and thinned until they reached the desired electron transparency. The battery slice was analyzed with a convergent electron beam with controlled convergence semi-angles (2 mrad and 0.15 mrad) to acquire a series of pixelated diffraction patterns. A high concentration of liquid electrolytes was used, consisting of 4.6 m lithium bis(fluorosulfonyl)amide (LiFSI) and 2.3 m lithium bis(trifluoromethanesulfonyl)imide (LiTFSI) in dimethoxymethane (DME). This is equivalent to a concentration of approximately 3 M LiFSI and 2 M LiTFSI. For the sake of simplicity, we will term this a "high concentration electrolyte". This liquid electrolyte is known to effectively prevent dendrite formation below a current density of 2 mAh/cm$^2$,[13–16] and we hypothesize that the structure of the SEI created by the decomposition of this liquid electrolyte is important to the overall performance.

**Figure 1b** displays a cryo-lift-out lamella with lithium grains plated at 0.5 mA/cm$^2$ to a capacity of 1 mAh/cm$^2$. Their grain size spans several micrometers in width. This morphology differs from the dendritic structures typically observed with other liquid electrolytes[17,18]. Our regions of interest in this study are the SEIs existing at the bottom and top marked by a red and black box in **Figure 1b**. A subset of the collected diffraction patterns (convergence semi-angle of 2 mrad) is shown in **Figure 1c**, which illustrates the nature of the data obtained at the interface between the Cu current collector and a lithium grain. The diffraction patterns can be classified into the three types based on their characteristic diffraction patterns: single crystalline lithium metal, an SEI domain, and a single crystalline Cu grain. Hence, these unique diffraction patterns at each pixel allow us to understand the structure of SEIs at the interfaces. It is noteworthy that the diffraction from the lithium region additionally shows Li$_2$O diffraction signal. This is a typical artifact due to oxidation of the air-sensitive metal, created by residual oxygen gas in the FIB chamber [19].

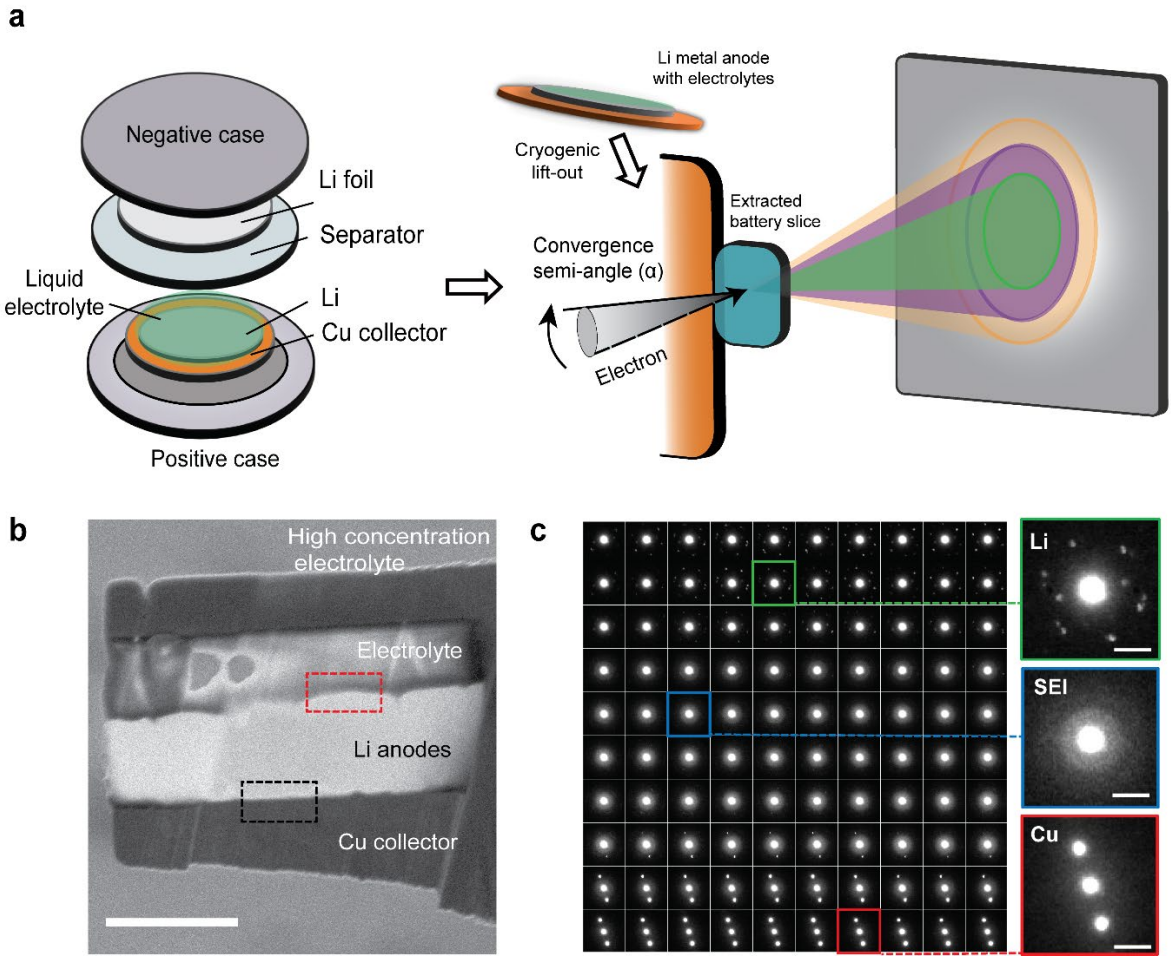

**Figure 1. Diffraction study of battery interfaces by cryo-SEND**. *(a) A schematic including creation of a battery slice and the SEND measurement approach. (b) Electron transparent cryo-lift-out lamella of lithium metal deposition with a high concentration electrolyte (a mixture of 4.6 m LiFSI and 2.3 m LiTFSI in DME) imaged using secondary electrons in the scanning electron microscope (SEM) column. (c) A subset of the diffraction patterns (10 × 10 pixels) from a total of 90 × 60 SEND data acquired from the bottom region of the lamella, which is a black box region in (b). The display of the diffraction patterns was binned by two to increase visibility of the individual patterns. The cropped diffraction patterns outlined by green, sky blue, and red lines illustrate three representative regions: a lithium grain, an SEI domain, and a Cu grain, respectively. Scale bars in (b) and (c) are 2 μm and 5 nm$^{-1}$.*

**Figure 2a** and **2b** show high-angle annular dark-field scanning transmission electron microscopy (HAADF) images of the bottom and the top region to indicate where the cryo-SEND measurement was performed. The bottom area has an approximately 50-60 nm thick SEI layer between a lithium metal grain and multiple grains of a Cu current collector, whereas a thinner SEI layer (≈ 40 nm) was present at the top area at the interface between the vitrified electrolyte layer and the plated lithium. The SEI layers observed in our analysis were significantly thicker than what previous cryo-EM studies have reported[8,20]. We believe this discrepancy arises from continuous deformation and thickening of the SEI as Li metal grains grow, exposing fresh surfaces to the liquid electrolyte and triggering the formation of additional SEI layers. Consequently, the SEI extracted using our cryo-FIB methodology likely represents the SEI at full capacity[7]. In contrast, SEIs prepared by other methods (e.g., deposition on Cu TEM grids[21]) may not fully capture the actual thickness present under operating conditions. The pixelated diffraction patterns collected at a convergence semi-angle of 2 mrad resulted in identification of amorphous structures in the SEIs, some of which are displayed below the HAADF images.

By analyzing the radially averaged diffraction intensity from the SEI region, we identified two distinct SEI types at the bottom interface between the lithium and the copper collector: we will label these as SEI Type 1 and SEI Type 2 (**Figure 2c**). To enhance the diffraction signal, the diffraction intensity was multiplied by scattering vector (Q), a method commonly employed in the form of $I \times Q^{(n)}$ (where n is an integer) in studies of X-ray and electron scattering[12,22]. The enhanced diffraction intensity ($I \times Q$, n = 1) from SEI Type 1 exhibits isotropic diffraction rings, suggesting the presence of short-range order in an amorphous structure. The first and second diffraction peaks in SEI Type 1 are centered at $\approx 0.33$ Å$^{-1}$ and 0.6 Å$^{-1}$, respectively, in reciprocal space, a spacing that is distinctly different than would be expected from compounds such as $Li_2O$ or LiF. The first peak of the short-range ordering in SEI Type 1 varies between 0.325 ~ 0.345 Å$^{-1}$ at each position, as illustrated by the four exemplary diffraction curves taken from this region (**Figure 2c**).

On the other hand, SEI Type 2 demonstrates three distinct localized diffraction intensities, with an exemplary diffraction curve displayed in **Figure 2c**. The first isotropic diffraction pattern demonstrates short-range ordering, which peaks at $\approx 0.3$ Å$^{-1}$, and which is not associated with the subsequent anisotropic second and third diffraction peaks occurring at $\approx 0.5$ Å$^{-1}$ and 0.7 Å$^{-1}$, respectively. The anisotropic contributions to the radially-averaged pattern are indicated by white dotted circles in **Figure 2a**, and their reflection orientations change in each diffraction pattern collected at a 4 nm in step size, indicating that they originate from local nanocrystalline regions which have different orientations with respect to the incident electron beam. The second and third diffraction peaks in the SEI Type 2 can be identified as originating from defective LiF crystals. This was further confirmed by collecting another SEND measurement at a lower convergence semi-angle (0.15 mrad). We found a noticeably weak (111) LiF reflection, possibly due to the presence of vacancies and/or atomic displacements in the fluorine lattice. (**Figure S1a-e**).

SEND measurements indicate that the SEI present at the interface between the frozen electrolyte and the lithium grain at the top of the sample was solely comprised of SEI Type 1 (**Figure 2b** and **2d**). An example diffraction pattern from the top region, represented by an orange curve in **Figure 2d**, has diffraction signals that match that of the SEI Type 1 observed at the bottom interface (with peaks at $\approx 0.33$ Å$^{-1}$ and 0.6 Å$^{-1}$). The structural ordering from the vitrified liquid electrolyte resulted in a short-range order with a peak at $\approx 0.285$ Å$^{-1}$ (navy curve in **Figure 2d**). This ordering is likely due to the aggregates formed by anion backbones (TFSI- and FSI-) present in the liquid electrolyte, as discussed in literature[23,24].

However, unlike the SEI at the region between the lithium and the copper current collector, we found a less dense SEI layer at the interface between the electrolyte and the lithium grain. leading to larger variations in short-range order. For example, **Figure 2d** displays diffraction curves collected from four regions between the Li grain and the electrolyte. The maximum scattering peak positions shifted from $\approx 0.34$ to 0.295 Å$^{-1}$ as the probe moved further away from the lithium grain and into the bulk of the vitrified electrolyte (**Figure 2d** and **Figure S2a-c**). This gradual shift to lower scattering vectors suggests that the SEI closer to the Li grain is denser, while the SEI closer to the bulk electrolyte is less compact. The uneven contour of the SEI and the electrolyte caused the electron beam to measure both components along the beam axis, superimposing the diffraction intensities from the two moieties.

These variations in ordering in both regions are shown in **Figure 2e** and **2f**, which maps the position of the maximum scattering peak in the two regions. The histograms – with their associated Gaussian distribution curves shown as dotted lines – indicate the presence of two distinct structures: SEI Type 1 and SEI Type 2. These structures have average scattering positions of 0.33 and 0.295 Å$^{-1}$, respectively, at the interface between the lithium and the copper current collector. While the average scattering position at the interface between the lithium and the electrolyte is in a similar range ($\approx$ 0.326 Å$^{-1}$), the histogram for the SEI near the copper foil shows a narrow peak (standard deviation of 0.019), while the interface near the electrolyte has a broader distribution (standard deviation of 0.026) due to the presence of an uneven interface along the electron beam direction. The electrolyte has an average peak position of 0.285 Å$^{-1}$.

Mapping the maximum peak positions of the short-range ordered structures suggests a layered structure of SEI Type 1 ($\approx$ 15 nm) and SEI Type 2 ($\approx$ 35 nm) at the interface between the lithium and the copper collector (**Figure 2g**). In contrast, the SEI region at the interface between the frozen electrolyte and the lithium has SEI Type 1 uniformly distributed across the Li grain with the outer SEI region overlapping with the electrolyte (**Figure 2h**). The presence of the two distinct layered SEI structures at the bottom suggests that structurally different layers of SEI form when Li nuclei are plated. For instance, the layered structure may indicate that SEI Type 2 forms in this less reductive environment before the formation of the Li nuclei; below 0 V, lithium nuclei start to form and grow. Simultaneously, the electrolytes are further reduced at the surface of Li, which is a more reducing environment ($\approx$ 0 V vs Li/Li$^+$). We hypothesize that the preferential formation of SEI Type 1, adjacent to lithium metal, is attributed to SEI decomposition products in this more reducing environment. This hypothesis is supported by the observation of SEI Type 1 between two lithium grains (**Figure S3a-e**) and SEI Type 2 as a decomposition product at a controlled voltage higher than Li/Li$^+$ (**Figure S4a-f**).

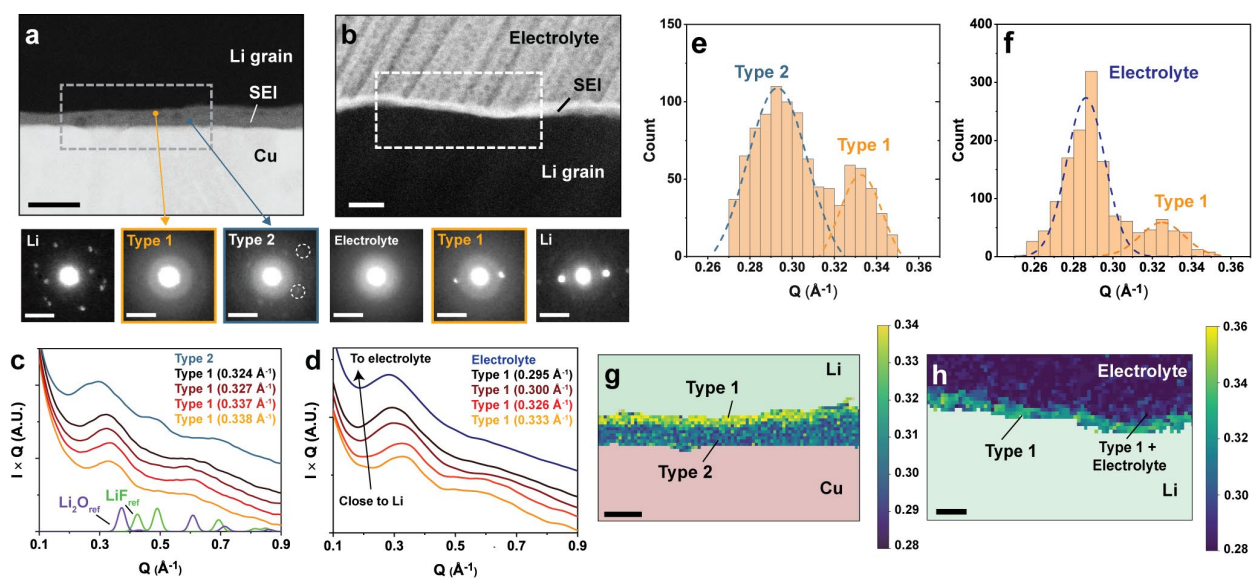

**Figure 2. SEI structural data.** *(a and b) HAADF-STEM images of the SEI at (a) the interface between the lithium deposits and the copper collector and (b) the interface between the frozen*

electrolyte and the lithium depositions. The display of diffraction patterns shows that the lower region is composed of two different structures (SEI Type 1 and SEI Type 2), while the upper region is comprised of only SEI Type 1. (*c* and *d*) Radial integration of diffraction intensity from the SEI Type 1, Type 2, and the electrolyte after signal enhancement. (*e* and *f*) Histograms of short-range ordering positions from (*e*) the lower lithium/copper collector interface and (*f*) the upper electrolyte/lithium interface. (*g* and *h*) Distributions of scattering peak positions of the short-range ordering among SEI Type 1, 2 and the electrolyte in the boxed regions at the lower SEI region (*g*) and the top SEI region (*h*). Map of the maximum peak positions of the short-range ordered structures in SEI Type 1 and Type 2 shows that a layered structure exists at the lithium/copper collector interface while uniform distribution of SEI Type 1 exists at the electrolyte/lithium interface. Scale bars in images (*a*), (*b*), (*g*) and (*h*) are 200 nm, 200 nm, 50 nm, and 100 nm, respectively. Scale bars for the diffraction patterns in (*a* and *b*) are each 0.5 Å$^{-1}$.

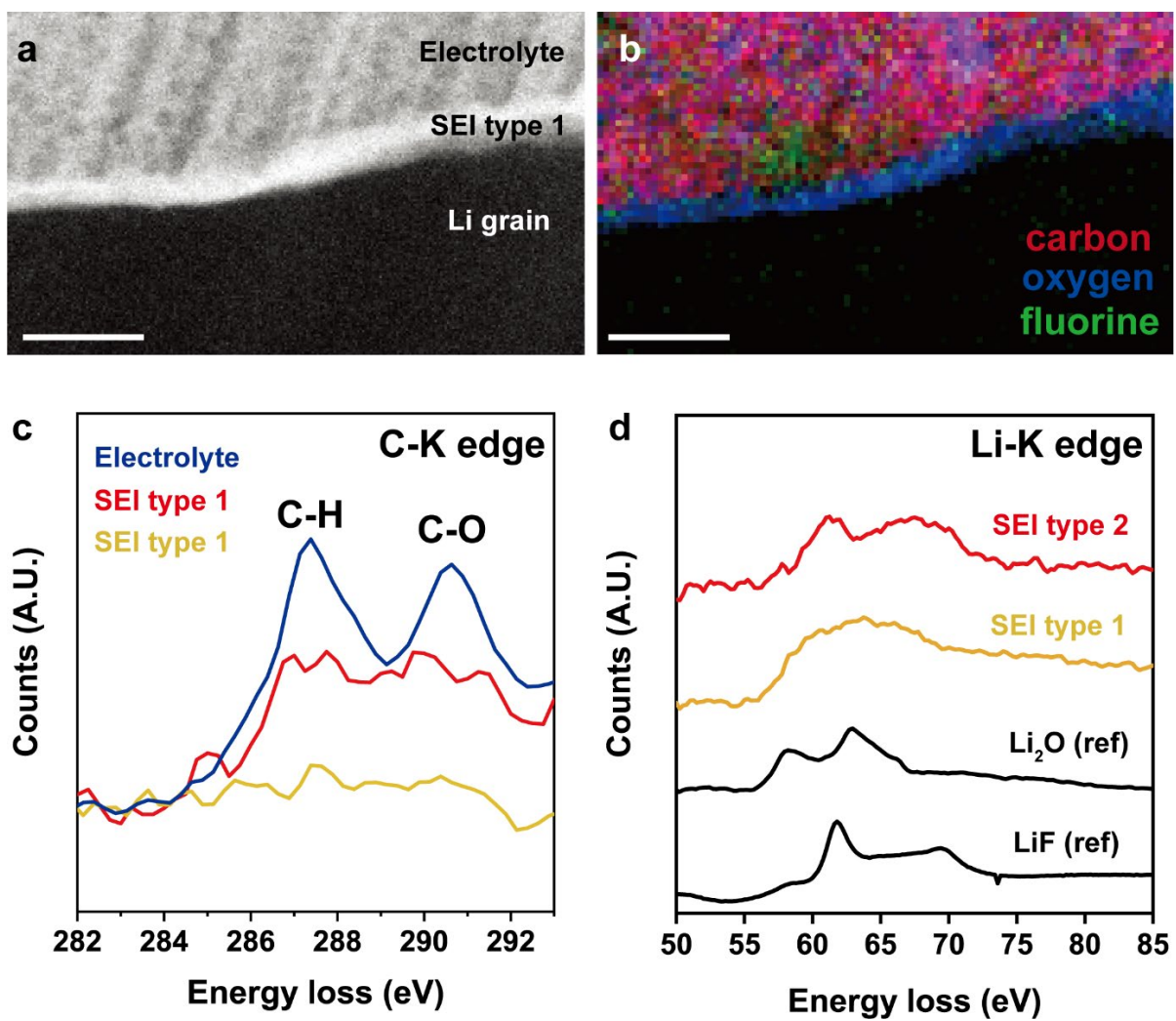

*Figure 3. Electron energy loss spectroscopy (EELS) from the SEIs.* (*a*) *A HAADF-STEM image showing the electrolyte, SEI Type 1, and a Li grain at the upper electrolyte/lithium interface.* (*b*) *Overlayed C, O, and F maps of the region in* (*a*). (*c*) *The fine structure of the C-K edge from the*

*electrolyte, SEI Type 1, and SEI Type 2. (d) Li-K edge of SEI Type 1 and SEI Type 2, with associated reference spectra. Scale bars in (a) and (b) are 200 nm.*

Elemental analysis by cryogenic energy loss spectroscopy (EELS) and energy dispersive spectroscopy (EDS) suggests that the origin of the short-range ordering in SEI Type 1 arises from inorganic bonding between Li, O, F, and S, and that it is essentially free of carbon (**Figure 3a**, **b** and **Figure S5a-e**). The carbon-free SEI results from preferential anion decomposition as the solvation of lithium ions changes when the lithium salts concentration becomes high[25], thus making the SEI rich in inorganic components. In contrast, the SEI Type 2 is chemically different, as confirmed by unique energy loss fine structures of O, F, and C, which are distinctly different than the spectra obtained from SEI Type 1 (**Figure 3c** and **Figure S5a-e**). Furthermore, the amorphous structure in SEI Type 1 exhibits a diffusive Li K-edge fine structure, which differs from the structures of crystalline lithium inorganic compounds such as $Li_2O$ and LiF (**Figure 3d**). On the other hand, the spectrum of SEI Type 2 indicates that LiF likely exists, which is consistent with the diffraction results. These distinct spectra between SEI Type 1 and 2, along with the diffraction results, indicates that they are chemically and structurally distinct.

We believe that the degree of order-disorder in the structure of SEIs plays a significant role in determining their ionic conductivity. Disordered states, such as short-range order, have been often reported to enhance diffusion of ion carriers compared to fully crystalline structures. For example, the short-range ordering of polyanion structures in borohydride families of solid electrolytes enables super ionic conductivity in both Na and Li ion systems[26]. This enhanced ion conductivity is associated with the order-disorder transition between crystalline structures and disordered symmetry phases, which promote the mobility of cations, by making them loosely bound to anions[27,28]. In addition, studies have shown that, such disorder state benefits cation diffusion as it is theoretically isotropic and does not rely on specific grain orientation for efficient cation conduction[29]. In contrast, crystalline ion conductors, unless they are single crystals, tend to form pronounced grain boundaries, which substantially leads to high interfacial barriers, thereby causing sluggish and uneven Li-ion diffusion[30]. The uniform structural ordering across the SEI, corroborated by SEND and EELS mapping, is expected to provide homogenized ion diffusion pathways, likely through interstitial or vacancy diffusion, thereby promoting Li diffusion and minimizing the formation of isolated Li particles. Therefore, the growth of the large granular Li grains is likely influenced by the enhanced and uniform conductivity of the short-range ordered structure.

Complete disorder may not be desirable for forming stable SEIs because those with a high degree of disorder are susceptible to creep due to their low elastic modulus. This results in SEIs failing to encapsulate Li deposits during subsequent cycles of dynamic volume expansion of Li metal anodes. We propose that some degree of ordering is essential for SEIs to stably function as protective layers. The short-range ordering from the high concentration electrolyte is expected to increase elastic modulus, likely offering advantages by being robust enough to accommodate changes associated with volume expansion/shrinkage of lithium anodes[1,8]. Enhanced elastic modulus due to the presence of short-range ordering has also reported in glassy electrolytes[31,32], solid-electrolyte polymers[33,34], and glassy inorganics[35–37]. Therefore, we hypothesize that the presence of the short-range ordering in the amorphous SEI layer is a key indicator of SEIs being stable and effective for high performance Li metal anodes.

As a test of our hypothesis, we compared the SEIs from additional two other electrolytes, 1 M $LiPF_6$ in EC/DEC and 1 M LiFSI in DME, which have been used in Li-metal batteries[38,39]. In contrast to the high concentration electrolyte, these electrolytes tend to deposit lithium dendrites, resulting in poor coulombic efficiencies[40,41]. To take diffraction patterns of the SEIs of the two electrolytes, two cryo-lift-outs were prepared and **Figure 4a** shows HAADF-STEM images of the respective lamellae and the diffraction results from the SEIs, demonstrating that the two electrolytes both preferentially deposit dendritic lithium morphologies. We did not see salt precipitation or solidification of liquid electrolytes during the sample preparation, based on the diffraction of the electrolytes (**Figure S6a** and **6b**).

When comparing these SEI diffraction patterns with that of the high concentration electrolyte, the presence of short-range ordering in the SEI is highly correlated to battery performance (**Figure 4b** and **4c**). For instance, the SEI of 1 M LiFSI in DME and 1 M $LiPF_6$ in EC/DEC did not have as strong short-range ordering as the SEI from the high concentration electrolyte. The average coulombic efficiencies of the batteries were calculated to allow correlation of the SEI structures with battery performances. Two methods were used to calculate the efficiencies: the standard cycle method (plating and stripping at 0.5 $mA/cm^2$ to 1 $mAh/cm^2$ of capacity) and the Aurbach method[42] to minimize effects of Cu current collectors (See methods and **Figure S7**). The high concentration electrolyte has the highest coulombic efficiency among these electrolytes (96.3% in the standard method and 98.2% in Aurbach method) while the moderate concentration electrolytes exhibit lower efficiencies measured in both methods. Overall, the ordering in the SEI structures is correlated to the increase in Coulombic efficiency among these electrolytes, indicating the short-range ordering is a key determinant in the performance of SEIs.

Finally, we attribute the observation of amorphous structures to the low dose capability of SEND. Our results show that the SEI Type 1 is extremely electron beam sensitive (**Figure S8**). After the total accumulate dose is around 3600 $e^-/Å^2$, we found that the short-range order was destroyed, creating a crystalline structure that corresponds to the diffraction from $Li_2O$. We found this not only in our diffraction studies, but also during the acquisition of EELS spectra, where doses greater than 400 $e^-/Å^2$ resulted in observable changes in the fine structure of the lithium and oxygen K-edge. This suggests that earlier SEI observations through phase-contrast imaging could have been under influence of beam artifacts, creating electron beam-induced artifacts.[10]

The combination of cryo-SEND and cryo-EELS allowed for the determination of both the structural and chemical composition of the SEI, providing a comprehensive understanding of its structure. In conclusion, this study showcases the power of cryo-STEM techniques in elucidating the structure-property relationships of SEIs in Li-metal batteries. The discovery of short-range ordering in high-performance SEIs and its correlation with battery performance provides valuable insights for the development of advanced battery systems.

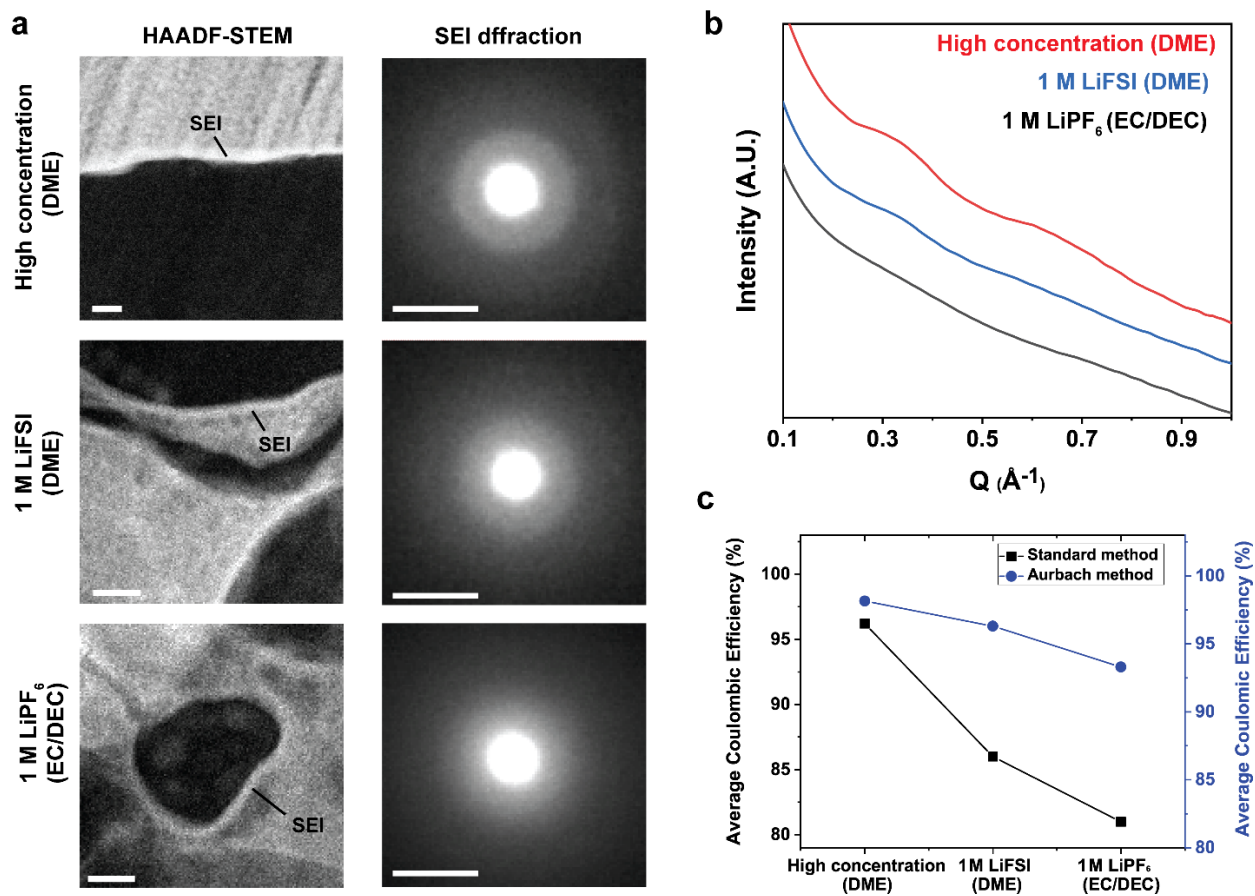

**Figure 4. Study of structural order present in the SEIs from liquid electrolytes.** *(a) HAADF-STEM images showing the lithium deposition morphology and associated SEIs from the three liquid electrolytes (High concentration in DME, 1 M LiFSI in DME, and 1 M LiPF$_6$ in EC/DEC) (left column). Representative diffraction patterns from their SEIs are displayed (right column). (b) Radial diffraction intensity of the three SEIs as a function of reciprocal space (Q). (c) Average Coulombic efficiencies of the three liquid electrolytes in Cu/Li cells, measured using two methods (Standard and Aurbach method). The scale bars in (a) are 100 nm for the HAADF-STEM image and 5 nm$^{-1}$ for the diffraction patterns, respectively.*

## Data availability

Data sets regarding scanning nanobeam electron diffraction and electron energy loss spectroscopy are available in DOI: (https://doi.org/10.5281/zenodo.13334386)

## Supporting Information

The Supporting Information includes additional experimental details, supporting data, and figures.

## Author Information

### Corresponding Author

Eric A. Stach
*Department of Materials Science & Engineering, University of Pennsylvania, Philadelphia, PA 19104, USA*
*Laboratory for Research on the Structure of Matter, University of Pennsylvania, Philadelphia, PA 19104, USA*

### Authors
Hyeongjun Koh
*Department of Materials Science & Engineering, University of Pennsylvania, Philadelphia, PA 19104, USA*

Eric Detsi
*Department of Materials Science & Engineering, University of Pennsylvania, Philadelphia, PA 19104, USA*

### Present Address
Hyeongjun Koh
*Andlinger Center for Energy and Environment, Princeton University, New Jersey, NJ 08540, USA*

### Author contributions

H.K. and E.A.S conceived the experiments. H.K. made Li anode samples and performed cryo-FIB/SEM and cryo-(S)TEM. H.K. processed the SEND data. H.K., E.D., and E.A.S. co-wrote the manuscript. E.A.S. oversaw this project.

### Notes

Authors declare that they have no competing interests.

### Acknowledgments

The authors gratefully acknowledge primary financial support from the National Science Foundation (NSF), Division of Materials Research (DMR), Future Manufacturing Research Grant #2134715. This work was carried out in part at the Singh Center for Nanotechnology, which is supported by the NSF National Nanotechnology Coordinated Infrastructure Program under grant NNCI-2025608. Additional support for the NSF through the University of Pennsylvania Materials Research Science and Engineering Center (MRSEC) (DMR-1720530; DMR-2309043). The authors show gratitude to Drs. Jamie Ford and Douglas Yates for support of Singh Center's

Nanoscale Characterization Facility and Prof. Paul Voyles for his comments. Finally, the authors also acknowledge the open access to Py4DSTEM, which made the analysis possible.